\documentclass[12pt,pra,aps,amssymb,amsmath,tightenlines,showpacs]{revtex4}
\usepackage{graphicx}

\newcommand{\half}{\mbox{$\textstyle \frac{1}{2}$}}

\newcommand{\ri}{\mbox{$\rm i$}}
\newcommand{\rd}{\mbox{$\rm d$}}
\newcommand{\bea}{\begin{eqnarray}}
\newcommand{\eea}{\end{eqnarray}}

\begin{document}

\title{Stochastic particle annihilation: a model of state
reduction in relativistic quantum field theory}

\author{Daniel~J.~Bedingham}
\email{d.bedingham@imperial.ac.uk}

\affiliation{Blackett Laboratory, Imperial College, London SW7
2BZ, UK }

\date{\today}

\begin{abstract}
A model of state reduction in relativistic quantum field theory
involving a nonlinear stochastic extension of Schr\"odinger's
equation is outlined. The eigenstates of the annihilation operator
are chosen as the preferred basis onto which reduction occurs.
These are the coherent states which saturate the bound of the
Heisenberg uncertainty relation, exhibiting classical-like
behavior. The quantum harmonic oscillator is studied in detail
before generalizing to relativistic scalar quantum field theory.
The infinite rates of increase in energy density which have
plagued recent relativistic proposals of dynamical state reduction
are absent in this model. This is because the state evolution
equation does not drive particle creation from the vacuum. The
model requires the specification of a preferred sequence of
space-like hyper-surfaces supporting the time-like state
evolution. However, it is shown that the choice of preferred surfaces
has no effect on perturbative results to second order in the
coupling parameter. It is
demonstrated how state reduction to a charge density basis can be
induced in fermionic matter via an appropriate coupling to a
bosonic field undergoing this mechanism.

\end{abstract}


\maketitle

\section{Introduction}
Much of the peculiar behavior associated with quantum
physics results from the fact that, although a quantum system can be in a
superposition of different states, whenever we make
measurements involving macroscopic apparatus,
a definite state is always registered.
The transition from a superposition
to a definite state is not described by Schr\"odinger's
equation. How then, if the constituents of the apparatus are also
described by Schr\"odinger's equation, does this quantum state
reduction come about?

Stochastic generalizations of Schr\"odinger's equation have been
proposed by a number of authors in answer to the problem of measurement \cite{pearlorig,gisin,ghir3,dios2,ghir2}
(for a review see \cite{Bass,Pear2}).
The key idea is that measurement is understood
as the realization of a random
process in the Hilbert space of state vectors where unwanted superpositions of
states are unstable.
The appeal of these models rests
on two fundamental properties: (i) they reproduce quantum effects on small
scales with negligible modification to standard quantum theory,
and (ii) they lead to rapid, objective state vector collapse on large scales
with probabilities given by the laws of standard quantum
mechanics. The result is that superpositions of states for macroscopic objects
are suppressed whilst individual particles continue to behave
according to quantum theory.

The usual approach is to substitute Schr\"odinger's equation with
a quantum state diffusion equation of the form
\bea
\rd|\phi_t\rangle = \left(C \rd t
+ {\bf A} \cdot \rd {\bf X}_t \right)
|\phi_t\rangle.
\eea
Here $\{{\bf X}_t\}$ is a (vector-valued) It\^o process and ${\bf A}$, $C$ are operators
(the Schr\"odinger equation can be recovered by setting $C=-\ri H$ and ${\bf A}=0$).
With appropriate choices for the drift and volatility of $\{{\bf X}_t\}$ the quantum state
typically evolves into an eigenstate of the
operator ${\bf A}$. The choice of ${\bf A}$ leads to a preferred
basis. In the quantum mechanical case, the standard choice is
a locally averaged position state basis in
order to reproduce the definite localization of objects at the classical scale.
Another idea is to use an energy state basis \cite{lane,Dorj2,adle2}.
These models have the desirable property that energy is conserved in expectation.
A general solution to the energy-based state diffusion with time-dependent coupling
has recently been found \cite{Dorj}.

At present, non-relativistic proposals are seen to have
sufficiently negligible effects on the quantum scale in order to
be indistinguishable from standard quantum theory for current
experimental technologies \cite{adleparams}. At the same time
these proposals offer a consistent understanding of classical and
quantum domains. However, so far, relativistic field theoretic
formulations generally predict an infinite rate of particle
creation due to the coupling of a classical stochastic field to a
quantum scalar field \cite{ghir1,pear3,adle,Bass}. Some previous
attempts to resolve this problem have involved modifying the
stochastic field to prevent high-energy excitations
\cite{Pear2,Pear}, or coupling the noise source not locally to the
quantum field but to the integral of quantum fields over some
space-time region \cite{Nicr}.
A quantum mechanical model for a relativistic particle
has been developed in reference \cite{tumul} although
this model does not include interactions.

In this paper we outline an alternative proposal in which the
stochastic field is coupled only to the annihilation operators of
the quantum scalar field (via a local interaction term). The
scalar field cannot then be excited by the stochastic field. As a
consequence the infinite rates of energy increase are avoided.
Instead we see an expected energy loss to the stochastic field
which can be controlled to a negligibly small level by an
appropriate choice for the coupling parameter. A related idea has
been employed in reference \cite{bass2} to control energy increase in
models of non-relativistic state reduction.

We find that in order to construct a satisfactory model of state
reduction in relativistic quantum field theory we must assume a
preferred sequence of space-like hyper-surfaces supporting the
evolution of the quantum state. The reason is that the stochastic
field is coupled to local operators which do not commute at
space-like separation. The state evolution equation is therefore
path-dependent. The fixed sequence of space-like hyper-surfaces
constrains the evolution such that only one path is possible,
ensuring a well defined evolving state. We do not propose a rule
for how the surfaces are chosen and regard them as a hidden
property of the state.

Our state evolution equations are of relativistically
invariant form so that all observers will agree on
outcomes. However, the choice of surface is responsible
for identifying a preferentially selected local frame.
The idea that dynamical reduction models might break Lorentz invariance in this way
has been suggested before by Pearle \cite{pearshape}, who considered
a stochastic field coupled to a generalized mass-density field
which does not commute at space-like separation. There it was shown that
the commutator decays on a length scale corresponding to the particle's
Compton wavelength, providing a sense in which the model is
quasi-relativistic.

By performing perturbative calculations involving an expansion in the coupling
parameter we are able to quantify the effect of a particular choice
of the space-like hyper-surfaces. We find that the choice has no effect on the
lowest order expressions describing state reduction.
This offers an alternative way to understand the quasi-relativistic
nature of this type of model.

We will see that the quantum state evolves towards the
eigenstates of the annihilation operators. In quantum mechanics
these are well understood as coherent states (see e.g.\ ref.\ \cite{optics}). The coherent states
have long been regarded as a close quantum approximation to
idealized classical states and therefore constitute a natural choice
for the preferred basis states in a quantum state reduction model.

The paper is organized as follows. In section \ref{HO} we
demonstrate the state reduction mechanism for the simple case of a
quantum harmonic oscillator. By analyzing the quantum variance
processes we are able to demonstrate that state reduction occurs,
and to estimate the associated reduction timescale. We also
examine how the expectation of energy evolves and demonstrate that
initial quantum probabilities match with the probabilities of
stochastic outcomes in a simple example. We conclude the section
with some numerical results which confirm our analysis.

In section \ref{rell} we extend the formalism to a relativistic quantum scalar field.
We adopt the interaction picture of Tomonaga and Schwinger \cite{Tomo,Schw} to describe a state
defined on some space-like hyper-surface evolving in a time-like manner. Once we
have examined this picture in detail, we
proceed to demonstrate the reductive properties. We show how
this mechanism of state reduction for a bosonic field could induce a state reduction
to some charge state basis in a fermionic field.
We end in section \ref{conc} with some concluding remarks.

\section{Quantum mechanical harmonic oscillator}
\label{HO}
The device we shall use to represent quantum state reduction will be presented
for the case of $(0+1)$-dimensional scalar field theory, i.e. the quantum mechanical
harmonic oscillator.
The commutation relation between position and momentum is given by $[x,p] = \ri$.
We define creation and annihilation operators in the standard way as follows
\bea
\left\{ \begin{array}{l} a=\sqrt{\frac{\omega}{2}}(x+\ri p\omega^{-1} ) \\
a^{\dagger}=\sqrt{\frac{\omega}{2}}(x-\ri p\omega^{-1}) \end{array} \right.
\quad \Leftrightarrow \quad
\left\{ \begin{array}{l} x=\frac{1}{\sqrt{2\omega}} (a+a^{\dagger}) \\
p=-\ri\sqrt{\frac{\omega}{2}}(a-a^{\dagger}) \end{array} \right.
\eea
These operators satisfy the commutation relation $[a,a^{\dagger}]=1$.
The Hamiltonian for the harmonic oscillator is given by
\bea
H=\half p^2 +\half \omega^2 x^2=\omega (a^{\dagger}a+\half)=\omega (N+\half),
\eea
where $N=a^{\dagger}a$ is the particle number operator.
Units are chosen such that $\hbar=1$ for the sake of simplicity.

The Schr\"odinger equation expressed in differential form is
$\rd|\psi_t\rangle = -\ri H|\psi_t\rangle \rd t$.
We extend this in the following way
\bea
\rd|\psi_t\rangle = \left\{\left[-\ri H -\half\lambda^2(a^{\dagger}-\bar{a}_{t})a
+\half\lambda^2(a-\bar{a}_{t})\bar{a}_{t}\right]\rd t
+\lambda(a-\bar{a}_{t})\rd B_t \right\}
|\psi_t\rangle,
\label{sse1}
\eea
where
\bea
\bar{a}_{t} = \half\langle\psi_t|(a+a^{\dagger})|\psi_t\rangle,
\eea
and $\lambda$ is a constant parameter of dimension $[{time}]^{-1/2}$.
Denoting unconditional expectation with respect to the physical probability measure $\mathbb{P}$
by $\mathbb{E}^\mathbb{P}[\cdot]$, the differential $\rd B_t$ is an increment
of real $\mathbb{P}$-Brownian motion with the properties that
$\mathbb{E}^\mathbb{P}[\rd B_t]=0$, $(\rd B_t)^2 =\rd t$, and increments
at different times are independent.
Equation (\ref{sse1}) can be derived (see \cite{Bass}) by first assuming a state evolution equation
of the form $\rd|\phi_t\rangle = \left(C\rd t+\lambda a\rd X_t\right)|\phi_t\rangle$ where
$|\psi_t\rangle=|\phi_t\rangle \langle\phi_t |\phi_t\rangle^{-1/2}$ and where $\{X_t\}$ is
a $\mathbb{Q}$-Brownian motion. The physical measure $\mathbb{P}$ is related to $\mathbb{Q}$
through $\mathbb{P}(A)=\mathbb{E}^{\mathbb{P}}[\mathbf{1}_A]=\mathbb{E}^{\mathbb{Q}}[\langle\phi_t |\phi_t\rangle \mathbf{1}_A]$ for some
event $A$ measurable at time $t$, where $\mathbf{1}_A = 1$ if $A$ is true and 0 otherwise. This
choice of physical probability measure is the counterpart to the postulate of standard quantum
mechanics on the outcomes of measurement processes \cite{Bass}.

Note that since the state evolves according to equation
(\ref{sse1}) by the action of only the number operator and the
annihilation operator, a final state with higher energy than any
of those states contributing to the initial superposition
$|\psi_0\rangle$ cannot occur. This ensures that as long as the
initial state has finite energy, subsequent evolved states must
also have finite energy.

We proceed by demonstrating that equation (\ref{sse1}) preserves the norm of a
state. Denoting $|\rd\psi_t\rangle = \rd|\psi_t\rangle$ we have
\bea
\rd (\langle\psi_t|\psi_t\rangle) &=&\langle\rd\psi_t|\psi_t\rangle
+\langle\psi_t|\rd\psi_t\rangle
+\langle\rd\psi_t|\rd\psi_t\rangle  \nonumber\\
&=&\langle\psi_t|\left[\ri H
-\half\lambda^2a^{\dagger}(a-\bar{a}_{t})
+\half\lambda^2(a^{\dagger}-\bar{a}_{t})\bar{a}_{t}\right]|\psi_t\rangle
\rd t+\langle\psi_t|\lambda(a^{\dagger}-\bar{a}_{t})|\psi_t\rangle \rd B_t\nonumber\\
&&+\langle\psi_t|\left[-\ri H
-\half\lambda^2(a^{\dagger}-\bar{a}_{t})a
+\half\lambda^2(a-\bar{a}_{t})\bar{a}_{t}\right]|\psi_t\rangle \rd t
+\langle\psi_t|\lambda(a-\bar{a}_{t})|\psi_t\rangle \rd B_t\nonumber\\
&&+\langle\psi_t|\lambda^2(a^{\dagger}-\bar{a}_{t})(a-\bar{a}_{t})|\psi_t\rangle \rd t \nonumber\\
&=& 0.
\label{normal}
\eea
For convenience we take the norm of the initial state $|\psi_0\rangle$ to be unity.
Further, we make the following definitions
for the conditional expectation and conditional variance of some operator $O$ with respect to the
state $|\psi_t\rangle$ at time $t$
\bea
O_t = \langle\psi_t|O|\psi_t\rangle
\quad {\rm and} \quad
{V}^O_t =\langle \psi_t|(O^{\dagger}-O_t^*)(O-O_t)|\psi_t\rangle \nonumber,
\eea
and the conditional covariance of two operators $O$ and $O'$
\bea
{V}_t^{O,O'} =\langle \psi_t|(O^{\dagger}-O_t^*)(O'-O'_t)|\psi_t\rangle.\nonumber
\eea
In addition, we define the operator $\Delta O_t = O-O_t$.

Let us first consider the energy of the oscillator. It is straightforward to demonstrate that
the energy process $H_t = \langle\psi_t|H|\psi_t\rangle$ satisfies the evolution equation
\bea
\rd H_t
=-\lambda^2\omega N_t \rd t
+\lambda\omega \langle\psi_t| ( a^{\dagger}a^{\dagger}a+a^{\dagger}aa
-2a^{\dagger}a\bar{a}_{t})|\psi_t\rangle \rd B_t.
\eea
By integrating and taking the unconditional expectation we infer that
\bea
\mathbb{E}^{\mathbb{P}}[ H_t] =H_0-\lambda^2\mathbb{E}^{\mathbb{P}}\left[\int_0^t\rd u\omega N_u\right]
=H_0-\lambda^2\omega\int_0^t\rd u\mathbb{E}^{\mathbb{P}}\left[ N_u\right].
\label{eproc}
\eea
The second term on the right side is negative semi-definite.
Therefore, energy is lost from the harmonic oscillator on average at a
rate determined by $\lambda^2$. We demand that energy loss on a
macroscopic scale is negligible in order to conform with the energy conservation principle.
Taking the typical particle number in the state
$|\psi_t\rangle$ to be of order $N_0$, we therefore require that $\lambda^2 \omega N_0 \Delta t \ll H_0$
for typical timescales $\Delta t$. Equivalently we may say that $\lambda$ must be very small in
standard macroscopic units of time.
In this limit we have that $\mathbb{E}^{\mathbb{P}}[H_t]\simeq H_0$,
or that the expected energy is approximately conserved.
In addition, having very small $\lambda$ means that for a small number of particles, equation (\ref{sse1}) can be accurately
approximated by Schr\"odinger's equation.

\subsection{State reduction}

In order to see how the state reduction mechanism works we
consider the stochastic processes $a_t$ and ${V}^a_t$ for the
conditional expectation of the annihilation operator and the
associated conditional variance:
\bea
\rd a_t &=&-\ri \omega a_t
\rd t -\half\lambda^2 a_t \rd t
+\lambda\langle\psi_t|\left[(a+a^{\dagger})a-2\bar{a}_{t}a\right]|\psi_t\rangle
\rd B_t,
\label{aprocess}\\
\rd {V}^a_t &=& -\lambda^2 \left\{\langle\psi_t||\Delta
a_t|^2|\psi_t\rangle +
|\langle\psi_t|\left[(a+a^{\dagger})a-2\bar{a}_{t}a
\right]|\psi_t\rangle|^2\right\} \rd t \nonumber\\ && \hspace{1cm}
+\lambda \langle\psi_t| \left[(a^{\dagger}-\bar{a}_{t})|\Delta
a_t|^2 +|\Delta a_t|^2(a-\bar{a}_{t})\right]|\psi_t\rangle \rd
B_t. \label{vaprosses} \eea Integrating and taking the
unconditional expectation of equation (\ref{vaprosses}) we have
\bea \mathbb{E}^{\mathbb{P}}[ V^a_t] &=&V^a_0 -\lambda^2
\mathbb{E}^{\mathbb{P}}\left[\int_0^t\rd u V^a_u \right]
-\lambda^2 \mathbb{E}^{\mathbb{P}}\left[\int_0^t\rd u
|V^{(a+a^{\dagger}),a}_u|^2
\right]\nonumber\\
&=&V^a_0 -\lambda^2 \int_0^t\rd u\mathbb{E}^{\mathbb{P}}\left[ V^a_u \right]
-\lambda^2 \int_0^t\rd u\mathbb{E}^{\mathbb{P}}\left[ |V^{(a+a^{\dagger}),a}_u|^2
\right].
\label{expvp2}
\eea
Since the last two terms on the right side are positive
semi-definite, the unconditional expectation of the variance of $a$ cannot increase (i.e. $V^a_t$
is a super-martingale). If we suppose that these terms are
nonzero then $\mathbb{E}^{\mathbb{P}}[ V^a_t]\rightarrow 0$ for large times
and therefore $V^a_t\rightarrow 0$ i.e. the state approaches an $a$-eigenstate.
Otherwise, if for some time $t$ we have $\mathbb{E}^{\mathbb{P}}\left[ V^a_t
\right]=0$ and $\mathbb{E}^{\mathbb{P}}\left[|V^{(a+a^{\dagger}),a}_t|^2
\right]=0$, then $|\psi_t\rangle$ at that time must be an
$a$-eigenstate.
Note that the second of these two conditions is also satisfied when $|\psi_t\rangle$ is
a position eigenstate at time $t$. Since these are composed of an infinite number
of infinitesimal energy mode contributions,
we exclude this possibility.

In order to estimate the characteristic timescale for state reduction we approximate
equation (\ref{expvp2}) by freezing the stochastic terms on the right side
at $t=0$. In this approximation we find
\bea
\frac{\mathbb{E}^{\mathbb{P}}[ V^a_t] - V^a_0}{V^a_0} \simeq
-\lambda^2\left(1+ \frac{|V^{(a+a^{\dagger}),a}_0|^2}{V^a_0}\right)t.
\label{freeze}
\eea
Taking $V^a_0 \sim V^{(a+a^{\dagger}),a}_0  \sim {\cal O}(N_0)$
(corresponding, for example, to a superposition between a large excited state
and the vacuum state), the reduction timescale for the variance-decreasing process can
be estimated as
\bea
\tau_R \sim \frac{V^a_0}{\lambda^2|V^{(a+a^{\dagger}),a}_0|^2 }
\sim \frac{1}{\lambda^2 N_0 }.
\label{timescale}
\eea
This must be small in standard units for macroscopic objects such
that macroscopic superpositions are suppressed. For example, for
an oscillator with frequency of order $10^{14} {\rm s^{-1}}$
(corresponding to visible light), if we take $N_0=10^{23}$ and
$\hbar = 10^{-34}{\rm Js}$, then choosing $\lambda = 10^{-8}
s^{-1/2}$ would lead to energy loss at a rate of $10^{-13}{\rm
Js^{-1}}$ and state reduction on a timescale of order $10^{-7}{\rm
s}$. For one particle ($N_0=1$) energy loss is of order
$10^{-36}{\rm Js^{-1}}$ and the reduction timescale is
$10^{16}{\rm s}$ ($10^9$ yrs).

Once the system enters an $a$-eigenstate,
equation (\ref{aprocess}) reduces to
\bea
\rd a_t
=\left(-\ri \omega   -\half\lambda^2\right) a_t \rd t,
\eea
with solution $a_t = a_0\exp\{-\ri \omega t - \half\lambda^2 t\}$. The solution decays on timescale
$\lambda^{-2}$ which as stated earlier must be very large in standard macroscopic units of time.

So far we have demonstrated that our state evolution equation (\ref{sse1}) describes
state reduction to a coherent state on timescale $\tau_R$ given in equation (\ref{timescale}).
We have also shown that coherent states themselves
will decay to the vacuum state on a very long timescale $\lambda^{-2}$.
We conclude this subsection by
confirming that stochastic probabilities match with quantum probabilities
for the outcome of a simplified measurement.
Let us consider the projection operator of a particle number eigenstate $P_n=|n\rangle\langle n|$.
The conditional expectation of the projection operator
$P_{n,t} = \langle\psi_t|P_n|\psi_t\rangle$ obeys the evolution equation
\bea
\rd P_{n,t}=\lambda^2\big[(n+1)P_{n+1,t}-nP_{n,t}\big]\rd t
+\lambda\langle\psi_t| (a^{\dagger}P_n+P_n a-2\bar{a}_{t}P_n)|\psi_t\rangle \rd B_t,
\label{pprocess}
\eea
where the terms in square brackets on the right side corresponds to the background decay mechanism
occurring on timescale $\lambda^{-2}$. These terms together are small
when a given wavepacket is sufficiently smoothly varying in $n$.
(For example, a wavepacket centered at $n=n'$ with a standard deviation in $n$ of ${\cal O}(\sqrt{n'})$,
typically has $P_{n,t}\sim {\cal O}(1/\sqrt{n'})$
and $[P_{n+1,t}-P_{n,t}] \sim {\cal O}(1/n')$, resulting
in $[(n+1)P_{n+1,t}-nP_{n,t}]\sim {\cal O}(1)$. These
orders of magnitude correspond to a minimum uncertainty coherent state wavepacket.)

Consider now an initial superposition state $|\psi_0\rangle$
consisting of the vacuum
state $|0\rangle$ and some excited coherent state $|\alpha_{0}\rangle$.
Suppose further that $\langle 0|\alpha_t\rangle\simeq 0$.
We may think of this situation as corresponding to a superposition
of null and positive readings on some measuring device.

After some time $t$ where $\tau_R < t \ll \lambda^{-2}$ reduction has
occurred to a coherent state. This may be either the vacuum state
or $|\alpha_t\rangle$. The initial quantum probability for
registering the system in the vacuum state is $P_{{\rm vac},0} =
\langle\psi_0|P_{\rm vac}|\psi_0\rangle$ where $P_{\rm
vac}=|0\rangle\langle 0|$. From equation (\ref{pprocess}) we have
(upon ignoring the terms in square brackets)
\bea
\rd P_{{\rm vac},t} \simeq \lambda\langle\psi_t|( a^{\dagger}P_{\rm vac}
+P_{\rm vac} a -2\bar{a}_{t}P_{\rm vac})|\psi_t\rangle \rd B_t.
\eea
Now taking the unconditional expectation we have
\bea
P_{{\rm vac},0} \simeq \mathbb{E}^{\mathbb{P}} \left[P_{{\rm
vac},t}\right] \simeq
\mathbb{E}^{\mathbb{P}}\left[\mathbf{1}_{|\psi_t\rangle
=|0\rangle}\right].
\eea
The final approximation results
from the fact that the state at time $t$ is either the vacuum
state or the approximately orthogonal excited coherent state
$|\alpha_t\rangle$. This relation tells us that the initial
standard quantum estimate for the probability of finding the
system in the vacuum state is equal to the stochastic probability
of that outcome occurring in this model. The quantum and
stochastic probabilities for the other outcome must also be equal.

\subsection{Numerical simulations}

In order to confirm the reductive properties
we ran a numerical simulation of the quantum state evolution. We considered an
initial state corresponding to an equal superposition of two $a$-eigenstates
with eigenvalues 0 and 8 respectively.
We have set the parameters to $\lambda=0.5$ and $\omega = 1$.
This choice means we observe state reduction for small numbers of particles
with only a small degree of energy loss. Since $N_0\sim 32$
we estimate the reduction timescale by equation (\ref{timescale}) to be
$\tau_R\sim 0.125$. The decay timescale is given by $\lambda^{-2}\sim 4$.
These order-of-magnitude estimates are confirmed by figures 1 and 2
which show sample paths for the conditional expectation of energy and for the
conditional variance in $a$ respectively. We see that the state
evolves into either one of the two possible coherent states. One of these
states is the vacuum state, the other corresponds to the (slowly decaying)
non-vacuum coherent state.

In addition we have estimated the
probabilities of the two possible outcomes by running 100 sample paths.
We find probabilities of 0.47 for the vacuum state and 0.53 for the
non-vacuum state (the standard deviation of this estimate is 0.1).

\begin{figure}[t]
\label{fig1}
\begin{center}
\includegraphics[width=10cm]{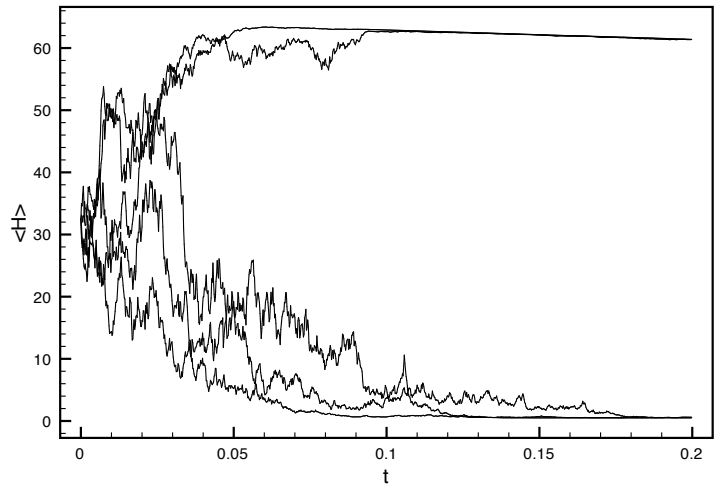}
\caption{ Conditional expectation of energy. The plot shows five
realized paths for an initial state corresponding to an equal
superposition of two coherent states with expected energies 0.5
and 64.5 respectively. In the cases where the state reduces to the
excited coherent state we note a slow decay in energy. This is
expected to occur on a timescale of order $\lambda^{-2}\sim 4$ in
this example ( $\lambda=0.5$ and $\omega = 1$). }
\end{center}
\end{figure}

\begin{figure}[t]
\label{fig2}
\begin{center}
\includegraphics[width=10cm]{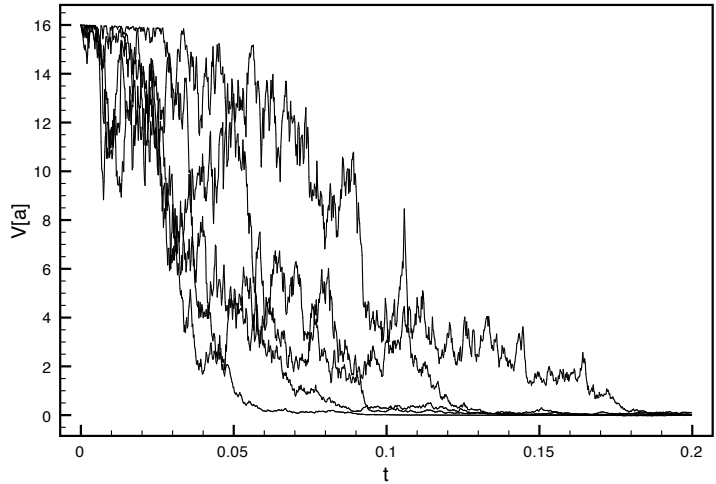}
\caption{
Conditional variance of the annihilation operator. The sample paths correspond to
those in figure 1 ( $\lambda=0.5$ and $\omega = 1$).
}
\end{center}
\end{figure}

\section{Relativistic quantum field theory}
\label{rell}
Here we generalize the analysis of the previous section to the case of relativistic quantum field theory.
(For a discussion of the conceptual issues surrounding the formulation of relativistic state reduction models,
see \cite{ghir1,Ghir4,Fay1,Fay2}.)
Given that experimental evidence conforms to the principle of relativistic invariance it is natural to
require this condition of our model. This has been
a longstanding problem in the field of dynamical state reduction models. The reason is that
while state reduction can be modelled easily enough, by coupling a stochastic process to a quantum field we
generate energy at an infinite rate.
We will resolve this issue by coupling only the annihilation operators of the quantum field
to the stochastic process (as in the case of the harmonic oscillator discussed in the
previous section). This will ensure that energy cannot be
created from the vacuum.


A natural formulation of relativistic quantum field theory for the consideration of an evolving state
is the one due to Tomonaga and Schwinger \cite{Tomo,Schw,pear3}.
We write the Hamiltonian density at space-time point $x$ in the form $H(x)=H_0(x)+H_{\rm int}(x)$,
where $H_0$ is the free field Hamiltonian and $H_{\rm int}$ is an interaction term.
Then evolution of the quantum state is described by the Tomonaga equation:
\bea
\ri\frac{\delta}{\delta\sigma(x)}|\Psi(\sigma)\rangle=H_{\rm int}(x)|\Psi(\sigma)\rangle.
\label{tomonaga}
\eea
The state is defined on some space-like three-surface $\sigma$, and functional
differentiation is defined with respect to some point $x$ lying on $\sigma$.
Given two space-like surfaces $\sigma$ and $\sigma'$ differing only by some
infinitesimal spacetime volume $\rd\omega_x$ at point $x$ (see figure 3) the functional
derivative can be expressed as
\bea
\frac{\delta|\Psi(\sigma)\rangle}{\delta\sigma(x)}=
\lim_{\sigma'\rightarrow\sigma}\frac{|\Psi(\sigma')\rangle-|\Psi(\sigma)\rangle}{\rd\omega_x}.
\eea
Equation (\ref{tomonaga}) describes the evolution of the quantum state in
terms of incremental time-like advancements of individual points on a
space-like surface. The operator $H_{\rm int}$ must be a scalar quantity
in order that equation (\ref{tomonaga}) has a relativistically invariant
form. In addition, the constraint $[H_{\rm int}(x),H_{\rm int}(x')]=0$
for space-like separated $x$ and $x'$ is imposed so that the ordering of points undergoing
time-like advancement is irrelevant. We will consider the possibility of a
definite ordering of all space-time points, allowing us to break the commutation
constraint.

In differential form the Tomonaga equation
can be represented as follows
\bea
\rd_x|\Psi(\sigma)\rangle =-\ri H_{\rm int}(x)|\Psi(\sigma)\rangle\rd\omega_x.
\eea
We proceed by generalizing this equation to a diffusion equation.

\begin{figure}[t]
\label{fig3}
\begin{center}
\includegraphics[width=8cm]{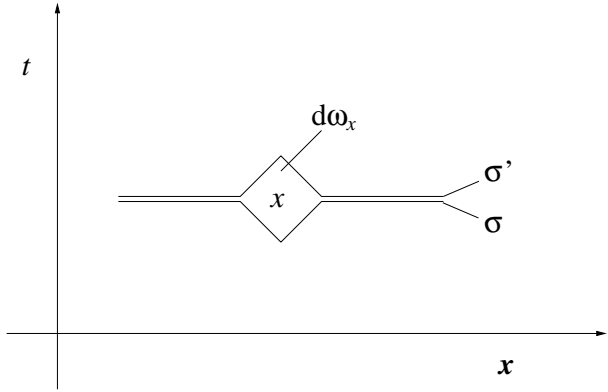}
\caption{Evolution between space-like hyper-surfaces $\sigma$ and $\sigma'$.}
\end{center}
\end{figure}

\subsection{Field state diffusion equation}

Previous approaches to modifying Schr\"odinger field dynamics have generally
involved the inclusion of a white-noise field term in the Tomonaga
equation (see e.g.\ \cite{Bass}).
Here we opt to formulate our model in terms of a Gaussian process.
We begin by defining $\rd W_{x}$ to be an increment of some real $\mathbb{Q}$-Brownian motion
with mean zero and covariance given by
\bea
\mathbb{E}^\mathbb{Q}[\rd W_{x}\rd W_{x'}] =  \delta_{x,x'} \rd\omega_x.
\eea
We may think of the Gaussian random variable $W(\sigma)$ defined on some surface $\sigma$ and
of $\rd W_{x}$ as the incremental difference in $W$ between two surfaces differing by some infinitesimal
space-time volume at point $x$.

We extend the differential Tomonaga equation to include a stochastic
term as follows:
\bea
\rd_x|\Phi(\sigma)\rangle =\big(-\half\lambda^2 A^{\dagger}(x)A(x)\rd\omega_x
+\lambda A(x)\rd W_x\big)|\Phi(\sigma)\rangle.
\label{sseA}
\eea
Here $A(x)$ is a scalar operator to be specified.

When using the Tomonaga picture, in order to set the initial conditions
we must specify an initial state on a definite initial space-like surface $\sigma_i$.
If we then wish to calculate the expected state at a later localized region in spacetime, we must specify
a final space-like surface $\sigma_f$ which includes that region.
To describe evolution from the initial state on
$\sigma_i$ to the final state on $\sigma_f$ we could choose
any causally ordered set of intermediate space-like surfaces (we write $\sigma'>\sigma$ if
$\sigma'$ is nowhere in the past of any point on $\sigma$). Each surface
will differ by only an incremental spacetime volume $\rd \omega_x$ from its neighboring
surfaces in the ordering.
If evolution of the state from the initial to the
final surface is independent of the ordering of intermediate surfaces we can say that
it is independent of any specific local frame. This is true of equation (\ref{sseA}) provided that
$[A(x),A(x')]=[A(x),A^{\dagger}(x')]=0$
for space-like separated $x$ and $x'$.

Assuming the usual rules of It\^o calculus we find
\bea
\rd_x(\langle\Phi(\sigma)|\Phi(\sigma)\rangle)
=2\lambda\langle{\bar{A}(x)}\rangle_{\sigma}\langle\Phi(\sigma)|\Phi(\sigma)\rangle\rd W_x,
\label{aeq}
\eea
where $\bar{A}(x)=\half(A(x)+A^{\dagger}(x))$,
$\langle\cdot\rangle_{\sigma}= \langle\Psi(\sigma)|\cdot|\Psi(\sigma)\rangle$,
and $|\Psi(\sigma)\rangle=|\Phi(\sigma)\rangle\langle\Phi(\sigma)|\Phi(\sigma)\rangle^{-\half}$
is the normalized state. The solution to this equation can be formally written as
\bea
\langle\Phi(\sigma_f)|\Phi(\sigma_f)\rangle &=&
\langle\Phi(\sigma_i)|\Phi(\sigma_i)\rangle+
2\lambda\int_{\sigma_i}^{\sigma_f}
\langle{\bar{A}(x)}\rangle_{\sigma}\langle\Phi(\sigma)|\Phi(\sigma)\rangle\rd W_x
\label{inner1}\\
&=&\langle\Phi(\sigma_i)|\Phi(\sigma_i)\rangle\exp\left\{
2\lambda\int_{\sigma_i}^{\sigma_f}\langle{\bar{A}(x)}\rangle_{\sigma}\rd W_x
-2\lambda^2\int_{\sigma_i}^{\sigma_f}\langle{\bar{A}(x)}\rangle^2_{\sigma}\rd \omega_x
\right\}.
\eea
We next introduce the physical measure $\mathbb{P}$ such that for
a random variable $X$, measurable on surface $\sigma_f$, the $\mathbb{P}$-expectation is given by
\bea
\mathbb{E}^{\mathbb{P}}[X] = \mathbb{E}^{\mathbb{Q}}
\left[\frac{\langle\Phi(\sigma_f)|\Phi(\sigma_f)\rangle}{\langle\Phi(\sigma_i)|\Phi(\sigma_i)
\rangle}X
\right].
\label{probrule}
\eea
The physical measure $\mathbb{P}$ assigns physical probabilities
to possible measurable outcomes.
We have from equation (\ref{inner1}) that $\mathbb{P}(\Omega)=\mathbb{E}^{\mathbb{P}}[1]=1$ as required of a probability measure.
Also, as a consistency check, given the tower law of $\mathbb{Q}$-expectation, it can be shown that
the tower law of $\mathbb{P}$-expectation also holds:
\bea
\mathbb{E}^{\mathbb{P}}[X]=\mathbb{E}^{\mathbb{P}}[\mathbb{E}^{\mathbb{P}}[X|\sigma]].
\eea
Here $\sigma$ is some surface such that $\sigma_f>\sigma>\sigma_i$, and
by conditioning on $\sigma$ we mean that all $\rd W_x$ to the past of $\sigma$ are known.
It therefore makes no difference for the final outcome if we condition on some
intermediate surface before taking the expectation at $\sigma_i$.
The application of equation (\ref{probrule}) therefore provides a consistent
way of assigning physical probabilities to outcomes. This allows us to describe state
evolution in terms of the $\mathbb{Q}$-Brownian motion
before using the $\mathbb{P}$-measure to determine physical probabilities at the end of the calculation.

We can also express the state evolution equation directly in terms of
a $\mathbb{P}$-Brownian motion as follows. First we choose a definite sequence of
space-like hyper-surfaces $\{\sigma\}$ (with $\sigma_f >\sigma >\sigma_i$) to support our state evolution. We then define the process
$B(\sigma)$ by the solution to the following stochastic equation
\bea
\rd B_x=\rd W_x-2\lambda\langle\bar{A}(x)\rangle_{\sigma}\rd \omega_x.
\eea
Here $\sigma$ is different from its succeeding surface only by some incremental space-time
volume about $x$. It can be shown that $\mathbb{E}^{\mathbb{P}}[\rd B_x] = 0$ and
$\mathbb{E}^{\mathbb{P}}[\rd B_x\rd B_{x'}]=\delta_{x,x'}\rd \omega_x$.
Therefore $\rd B_x$ is an increment of $\mathbb{P}$-Brownian motion.
Finally, writing equation (\ref{sseA}) in terms of
the normalized state $|\Psi(\sigma)\rangle$ and the $\mathbb{P}$-Brownian motion
$\rd B_x$ we find
\bea
\rd_x|\Psi(\sigma)\rangle =\big(\alpha(x,\sigma)\rd\omega_x
+\beta(x,\sigma)\rd B_x\big)|\Psi(\sigma)\rangle,
\label{SSE}
\eea
where (cf.\ equation (\ref{sse1}))
\bea
\alpha(x,\sigma) &=& -\half\lambda^2\big(A^{\dagger}(x)-\langle \bar{A}(x)\rangle_{\sigma}\big)A(x)
+\half\lambda^2\big(A(x)-\langle \bar{A}(x)\rangle_{\sigma}\big)\langle \bar{A}(x)\rangle_{\sigma},
\\
\beta(x,\sigma) &=& \lambda\big(A(x)-\langle \bar{A}(x)\rangle_{\sigma}\big).
\eea
In the case where $[A(x),A(x')]=[A(x),A^{\dagger}(x')]=0$ for space-like separated $x$ and $x'$,
this evolution equation must be independent of the choice $\{\sigma\}$ by construction.
On the other hand, if we allow for
$[A(x),A(x')]=[A(x),A^{\dagger}(x')]\neq 0$, the state evolution described by equation (\ref{sseA})
is $\{\sigma\}$-dependent and our choice of sequence $\{\sigma\}$ must be specific if
the model is to give unambiguous results.
Equation (\ref{SSE}) retains its
relativistically invariant form so that all observers will agree on outcomes.

We will be forced to choose operators $A(x)$ that do not commute at
space-like separation and therefore we must
specify a fixed sequence of evolving space-like hyper-surfaces.
This might seem a significant
compromise, however, it is not clear
that a freedom to choose any space-like surface is desirable in
a model of quantum state reduction.
Consider an entangled EPR pair
where one particle is measured at a space-like separation
from a region where we wish to consider the state of the other particle.
The state of the unmeasured particle depends on
whether the surface on which it is defined has the
measurement event in its past or future. Since we are free to choose this
surface, the state of the unmeasured particle is ambiguous.
As is argued in reference \cite{ghir1} this problem only persists for the
state reduction timescale so it can be ignored for macroscopic objects.
However, it is a difficulty if we intend for our state to represent the
microscopic world unambiguously.

If nature were to choose the specific sequence $\{\sigma\}$
this problem could be avoided. We
would have no freedom to choose the surface upon which the final state
is defined.
We do not suggest a rule for the choice. We only suggest that
relativistic invariance could be recovered in expectation by
assuming that future space-like surfaces are chosen at random from a uniform distribution
over the space of all future space-like surfaces.
Alternatively, we might simply be content to allow our model to break relativistic invariance
in its description of state reduction. For example,
the evolving surfaces could correspond to the constant time surfaces in the co-moving
frame of the Universe or to a local frame defined by the matter content of the state.

Without a rule for choosing the sequence of surfaces
we must quantify the effect of making different choices.
In fact, we can demonstrate that the imposed ordering of space-time points has negligible effect
in a perturbative calculation scheme involving the coupling parameter $\lambda$.
Given some operator $O$ such that $\rd_x O=0$ in the Tomonaga picture, we can use
(\ref{SSE}) to determine the dynamical equation satisfied by its conditional expectation as
\bea
\rd_x \langle {O} \rangle_{\sigma} =
\langle\alpha^{\dagger}(\sigma){O}+{O}\alpha(\sigma)+\beta^{\dagger}(\sigma){O}\beta(\sigma)\rangle_{\sigma} \rd \omega_x
+\langle\beta^{\dagger}(\sigma){O}+{O}\beta(\sigma)\rangle_{\sigma} \rd B_x,
\eea
(where the $x$ dependence of $\alpha$ and $\beta$ is assumed).
Integrating and taking the unconditional expectation we find
\bea
\mathbb{E}^{\mathbb{P}}\left[\langle {O} \rangle_{\sigma_f}\right] =
\langle {O} \rangle_{\sigma_i}
+\mathbb{E}^{\mathbb{P}}\left[
\int_{\sigma_i}^{\sigma_f}\langle\alpha^{\dagger}(\sigma){O}+{O}\alpha(\sigma)
+\beta^{\dagger}(\sigma){O}\beta(\sigma)\rangle_{\sigma} \rd \omega_x
\right].
\eea
Since $\alpha \sim {\cal O}(\lambda^2)$ and $\beta \sim {\cal O}(\lambda)$, we can expand
$\mathbb{E}^{\mathbb{P}}\left[\langle {O} \rangle_{\sigma_f}\right]$
perturbatively in $\lambda$ to second order
by freezing the stochastic state at the initial surface $\sigma_i$, that is,
\bea
\mathbb{E}^{\mathbb{P}}\left[\langle {O} \rangle_{\sigma_f}\right] \simeq
\langle {O} \rangle_{\sigma_i}
+\mathbb{E}^{\mathbb{P}}\left[
\int_{\sigma_i}^{\sigma_f}\langle\alpha^{\dagger}(\sigma_i){O}+{O}\alpha(\sigma_i)
+\beta^{\dagger}(\sigma_i){O}\beta(\sigma_i)\rangle_{\sigma_i} \rd \omega_x
\right].
\eea
In this approximation, even when the $A(x)$-operators do not commute at space-like separation,
the result only depends on the choice of initial and final surfaces, and not on
any ordering of space-time points within the integrated region.
The choice of intermediate surfaces will have no effect.
We will use equivalent frozen state approximations in
subsequent sections. The results will be Lorentz invariant in the sense outlined here.

We end this subsection by commenting on the ``Free Will Theorem"
\cite{FWT} which claims to show that relativistic dynamical
reduction models are incompatible with the experimenter's free
will to decide which observable to measure. In subsequent
responses \cite{fwt2,fwt3} it has been argued that the resolution
of this conflict can be found in nonlocality (see also
\cite{fwt4}). Certainly equation (\ref{SSE}) is explicitly
nonlocal through its dependence on the quantum state over the entire
space-like surface $\sigma$.
However, as pointed out by 't Hooft \cite{fwt5}, for
models of this type we should reconsider our notion of ``free
will''. For example, given some definite quantum state defined on
some initial surface $\sigma_i$, and given some realized $B(\sigma)$ for
every space-like surface $\sigma$ to the future of $\sigma_i$, then the
future quantum state is determined. This future
quantum state should describe all matter including the
experimenter's behavior. If we require free will in this
framework, it can only result from an inability to determine the
precise initial state \cite{fwt5}.

\subsection{Scalar field theory}

Having established the covariant form of the theory,
we now focus on a particular frame with space-like surfaces chosen
to be the constant time surfaces. We have
\bea
|\rd \Psi(t)\rangle = \int_{\bf x} \rd_{x}|\Psi(t)\rangle
= \int_{\bf x}\rd{\bf x}\big(\alpha(x)\rd t + \beta(x)\rd B_t({\bf x})\big)|\Psi(t)\rangle,
\label{evo}
\eea
with $\mathbb{E}^\mathbb{P}[\rd B_{t}({\bf x})\rd B_{t'}({\bf x'})] = \delta^3({\bf x-x'})\delta_{t,t'}\rd t$.
We use the integration subscript to avoid confusion over which variables are integrated over.
In this frame, time-independent operators in the Schr\"odinger picture are related to time-dependent
operators in the Tomonaga picture by the unitary transformation $O(t)= \exp\{\ri H_0 t\}O\exp\{-\ri H_0 t\}$,
where $H_0$ is the free field Hamiltonian.

We consider a real scalar field $\varphi$ defined in the Tomonaga picture by
\bea
\varphi(x)&=&\int\frac{\rd {\bf p}}{\sqrt{2\omega_{\bf p}}}
\left\{\exp{(\ri {\bf p}\cdot {\bf x}-\ri \omega_{\bf p} t)}a({\bf p})+
\exp{(-\ri {\bf p}\cdot {\bf x}+\ri \omega_{\bf p} t)}a^{\dagger}({\bf p})\right\},
\label{field}
\eea
with free Hamiltonian
\bea
H_0 &=& \int\rd {\bf x}\left\{\half\left(\partial_t\varphi(x)\right)^2
+\half\nabla\varphi(x)\cdot\nabla\varphi(x)
+\half m^2\varphi^2(x)\right\}\nonumber\\
&=&\int \rd{\bf p}\omega_{\bf p}\left\{a^{\dagger}({\bf
p})a({\bf p})+\half\delta^3(\bf 0)\right\},
\eea
where $\omega_{\bf p} =\sqrt{{\bf p}^2+m^2}$, and the creation and annihilation operators satisfy
the canonical commutation relations
$\lbrack a({\bf p}),a^{\dagger}({\bf p'}) \rbrack = \delta^3({\bf p-p'})$ and
$\lbrack a({\bf p}),a({\bf p'}) \rbrack = 0$.
The positive and negative frequency components of the field are given by
\bea
\varphi^{+}(x)&=&\int\frac{\rd {\bf p}}{\sqrt{2\omega_{\bf p}}}
\exp{(\ri {\bf p}\cdot {\bf x}-\ri \omega_{\bf p} t)}a({\bf p}),\\
\varphi^{-}(x)&=&\int\frac{\rd {\bf p}}{\sqrt{2\omega_{\bf p}}}
\exp{(-\ri {\bf p}\cdot {\bf x}+\ri \omega_{\bf p} t)}a^{\dagger}({\bf p}),
\eea
where $\varphi = \varphi^{+}+\varphi^{-}$. We define
\bea
\alpha &=&-\half\lambda^2\left(\varphi^{-}-\half\langle\varphi\rangle_t\right)\varphi^{+}
+\half\lambda^2\left(\varphi^{+}-\half\langle\varphi\rangle_t\right)\half\langle\varphi\rangle_t \\
\beta &=&\lambda\left(\varphi^{+}-\half\langle\varphi\rangle_t\right).
\label{alphbet}
\eea
Here $\langle\cdot\rangle_t=\langle \Psi(t)|\cdot|\Psi(t)\rangle$.
The constant parameter $\lambda$ in this model has dimension $[time]^{-1}$.
We ignore for now any other possible Hamiltonian interaction terms.
Since $\varphi^{+}$ and $\varphi^{-}$ do not commute, our choice of
constant-time surfaces must be considered special. Although, as we have seen
in the previous section, by using the frozen state approximation,
our results will be independent of any specific local frame.
\footnote{An alternative suggestion that we have explored is to remove the
on-shell constraint from the field creation and annihilation
operators. This enables us to construct local scalar field
operators which do commute at space-like separation. The hope is
that a state with only on-shell excitations might enforce the
on-shell condition. However, we have been unable to prevent
off-shell excitations from occurring (including faster-than-light
modes).}

In the same manner as (\ref{normal}) we can demonstrate that
\bea
\rd \langle \Psi(t)|\Psi(t)\rangle = 0,
\eea
so without loss of generality we may set $\langle \Psi(t)|\Psi(t)\rangle=1$ with the state
remaining normalized for all time.

Given some generic operator ${O}(t)$ in the Tomonaga picture, we may ask how its conditional expectation
evolves. We find (cf.\ \cite{adle})
\bea
\rd \langle {O} \rangle_t = \langle \rd {O}\rangle_t
+\int_{\bf x}\rd{\bf x}\langle\alpha^{\dagger}{O}+{O}\alpha+\beta^{\dagger}{O}\beta\rangle_t \rd t
+\int_{\bf x}\rd{\bf x}\langle\beta^{\dagger}{O}+{O}\beta\rangle_t \rd B_t(\bf x),
\label{expt}
\eea
where dependencies on spatial coordinates are understood. The first
term on the right side results from the standard unitary
evolution of the operator ${O}$ described by the
free Hamiltonian.

Similarly we can write an evolution equation for the conditional variance of an operator.
Recalling that $\Delta{O}_t={O}-\langle{O}\rangle_t$ and that the
conditional variance is given by $V_t[{O}] = \langle|\Delta{O}_t|^2\rangle_t$, we find
(again cf.\ \cite{adle})
\bea
\rd V_t[{O}] &=& \langle \rd{O}^{\dagger}\Delta{O}_t+\Delta{O}_t^{\dagger}\rd{O}\rangle_t
+\int_{\bf x}\rd{\bf x}\langle\alpha^{\dagger}|\Delta{O}_t|^2+|\Delta{O}_t|^2\alpha
+\beta^{\dagger}|\Delta{O}_t|^2\beta\rangle_t \rd t\nonumber\\
&&-\int_{\bf x}\rd{\bf x}\langle\beta^{\dagger}{O}^{\dagger}+{O}^{\dagger}\beta\rangle_t
\langle\beta^{\dagger}{O}+{O}\beta\rangle_t\rd t\nonumber\\
&&+\int_{\bf x}\rd{\bf x}\langle\beta^{\dagger}|\Delta{O}_t|^2+|\Delta{O}_t|^2\beta\rangle_t \rd B_t({\bf x}).
\label{vary}
\eea
Note that the third term on the right side of equation (\ref{vary}) is negative
semi-definite. This term is responsible for the variance reduction which we
can use to demonstrate state reduction (see next subsection).

We may apply equation (\ref{expt}) to the total
energy of the quantum field. Ignoring the vacuum energy and interactions, this is given by
\bea
H = \int \rd{\bf p}\omega_{\bf p}a^{\dagger}({\bf
p})a({\bf p}).
\eea
We find after some calculation that
\bea
\rd
\langle H \rangle_t  = -\half\lambda^2 \langle N\rangle_t\rd t
+\int_{\bf x}\rd {\bf x}
\langle\beta^{\dagger}H+H\beta\rangle_t \rd B_t({\bf x}),
\label{hproc}
\eea
where
\bea
 N = \int \rd{\bf p}a^{\dagger}({\bf
p})a({\bf p}).
\eea
Integrating and taking the unconditional expectation of the energy process at time $t$ we have
\bea
\mathbb{E}^\mathbb{P}[\langle H \rangle_t] = \langle H \rangle_0
-\half\lambda^2 \int_0^t \rd u \mathbb{E}^\mathbb{P}[\langle N\rangle_u],
\label{eproc2}
\eea
(cf.\ equation (\ref{eproc})). Since $\langle N\rangle_t$ is nonnegative,
it follows from (\ref{eproc2}) that
energy is lost on average as a result of
coupling the quantum field to a classical stochastic process.
However, the energy loss is finite and can be made negligible by
an appropriate choice of $\lambda$. This is to be contrasted with
some of the
previous attempts to construct a relativistic state reduction
model \cite{pear3,ghir1,adle}, where the energy density is seen to
increase at an infinite rate. The reason that we do not see an
infinite rate of energy density creation can be traced back to the
fact that the classical stochastic process is not coupled to the
particle creation operator and therefore cannot drive particle
creation from the vacuum.

Stochastic movements in the energy process will cease
when the quantum state is an eigenstate of the operator $\varphi^{+}$.
When this occurs, the final term on the right side of equation (\ref{hproc}) goes to
zero.

We can approximate equation (\ref{eproc2}) to ${\cal O}(\lambda^2)$ by freezing the
stochastic terms on the right side at time $t=0$. This gives
\bea
\mathbb{E}^\mathbb{P}[\langle H \rangle_t] \simeq \langle H \rangle_0
-\half\lambda^2 \langle N\rangle_0 t
\label{enzero}
\eea
This result depends on the initial state and on the integrated region of space-time
between the initial and final space-like hyper-surfaces. However, no ordering of space-time points is required.

\subsection{Quantum field state reduction}

To see the reductive properties we consider the particle annihilation operator $a({\bf p})$.
Using equation (\ref{expt}) we find
\bea
\rd \langle a({\bf p}) \rangle_t = -\ri\omega_{\bf p}\langle a({\bf p})\rangle_t\rd t
 -\frac{\lambda^2}{4\omega_{\bf p}} \langle a({\bf p})\rangle_t\rd t
+\lambda\int_{\bf x}\rd{\bf x}\langle\left(\varphi-\langle\varphi\rangle_t\right)a({\bf p})
\rangle_t \rd B_t({\bf x}).
\eea
Similarly using equation (\ref{vary}) and taking the unconditional expectation we have
\bea
\mathbb{E}^\mathbb{P}[ V_t[a({\bf p})]] &=&V_0[a({\bf p})] -\frac{\lambda^2}{2\omega_{\bf p}}
 \mathbb{E}^\mathbb{P}\left[\int_0^t\rd u V_u[a({\bf p})] \right]\nonumber\\
&&-\lambda^2 \mathbb{E}^\mathbb{P}\left[\int_0^t\rd u \int_{\bf x}\rd{\bf x}
|\langle\left(\varphi-\langle\varphi\rangle_t\right)a({\bf p})
\rangle_t|^2
\right]
\nonumber\\
&=&V_0[a({\bf p})] -\frac{\lambda^2}{2\omega_{\bf p}}
\int_0^t\rd u\mathbb{E}^\mathbb{P}\left[ V_u[a({\bf p})] \right]\nonumber\\
&&-\lambda^2 \int_0^t\rd u\mathbb{E}^\mathbb{P}\left[ \int_{\bf
x}\rd{\bf x}
|\langle\left(\varphi-\langle\varphi\rangle_t\right)a({\bf
p}) \rangle_t|^2 \right]. \label{energyvar} \eea Again we find
that the conditional variance for the annihilation operator is a
super-martingale. The expected variance decreases with time and
the quantum state evolves towards an eigenstate of the
annihilation operator. If we freeze the stochastic terms on the right side of
equation (\ref{energyvar}) we find
\bea
\mathbb{E}^\mathbb{P}[ V_t[a({\bf p})]]
\simeq V_0[a({\bf p})] -\frac{\lambda^2}{2\omega_{\bf p}}
 V_0[a({\bf p})] t -\lambda^2 \int_0^t\rd u\int_{\bf
x}\rd{\bf x}
|\langle\left(\varphi-\langle\varphi\rangle_0\right)a({\bf
p}) \rangle_0|^2.
\eea
Note that, as in equation (\ref{enzero}), the right side is independent of the
ordering of space-time points and therefore independent of the intermediate
space-like hyper-surfaces we have chosen to support our state evolution.
We may estimate the timescale for collapse
in the same manner as equations (\ref{freeze}) and
(\ref{timescale}) by taking $V_0[a({\bf p})]\sim N_0({\bf
p})=\langle a^{\dagger}({\bf p})a({\bf p})\rangle_0$ and
\bea
\int_{\bf x}\rd{\bf
x}|\langle\left(\varphi-\langle\varphi\rangle_t\right)a({\bf
p})\rangle_t|^2 \sim \int\rd {\bf p'} \frac{N_0({\bf p'}) N_0({\bf
p})}{2\omega_{\bf p'}},
\eea
from which we find
\bea \tau_R \sim
\frac{1}{\lambda^2  \int\rd {\bf p'} N_0({\bf p'})/(2\omega_{\bf
p'}) }.
\eea

As in the harmonic oscillator case, it is the third
term on the right side of equation (\ref{energyvar}) that leads to
variance reduction for macroscopic energy scales. The reduction
time is inversely proportional to the total number of excitations
in all modes. This will lead to rapid reduction for large scale
excitations. Each mode tends towards a coherent state. As this
occurs, we expect that the field tends towards classical behavior.

\subsection{Fermionic state reduction}

Here we introduce a fermionic field coupled to our proposed
scalar field theory in order to consider an
induced state reduction in the fermionic sector.
To see how this works let us set $\lambda$ to zero for now
and consider an interaction Hamiltonian of the type
\bea
H_{\rm int} (t) = \int_{\bf x}\rd {\bf x} j(x) \varphi(x).
\eea
Here $j$ is some Hermitian current operator associated with the fermionic matter field.
From equation (\ref{field}) we have
\bea
H_{\rm int} (t)&=&\int_{\bf x}\rd {\bf x} j(x)
\int\frac{\rd {\bf p}}{\sqrt{2\omega_{\bf p}}}\left\{\exp{(\ri {\bf p}\cdot {\bf x}-\ri \omega_{\bf p} t)}a({\bf p})+
\exp{(-\ri {\bf p}\cdot {\bf x}+\ri \omega_{\bf p} t)}a^{\dagger}({\bf p})\right\}\nonumber\\
&=&\int\frac{\rd {\bf p}}{\sqrt{2\omega_{\bf p}}}\left\{j^{\dagger}({\bf p},t)\exp{(-\ri \omega_{\bf p} t)}a({\bf p})+
j({\bf p},t)\exp{(\ri \omega_{\bf p} t)}a^{\dagger}({\bf p})\right\}.
\eea
Furthermore, we can formally solve the Tomonaga equation to find
\bea
|\Psi(t)\rangle = \exp{\left\{-\ri\int_0^t\rd u H_{\rm int}(u)\right\}}|\Psi(0)\rangle.
\eea
Now suppose that the fermionic state undergoes some spatial transfer of charge such that
a pulse of current occurs. If the fermionic state is a $j$-eigenstate, we
have
\bea
|\Psi(t)\rangle = \exp{\left\{\int\rd {\bf p}( \alpha({\bf p},t)a^{\dagger}({\bf p})-
\alpha^*({\bf p},t)a({\bf p}) ) \right\}}|\Psi(0)\rangle,
\eea
where the complex number $\alpha$ is given by
\bea
\alpha({\bf p},t) = -\ri\int_0^t\rd u \frac{j({\bf p},u)}{\sqrt{2\omega_{\bf p}}}\exp(\ri \omega_{\bf p} u),
\label{alpha}
\eea
and $j({\bf p},t)$ is the current eigenvalue at time $t$.
Using the commutation relations for the creation and annihilation
operators, and assuming that the initial $\varphi$ state is unexcited, we find
\bea
a({\bf p'})|\Psi(t)\rangle &=& a({\bf p'})\exp{\left\{\int\rd {\bf p}( \alpha({\bf p},t)a^{\dagger}({\bf p})-
\alpha^*({\bf p},t)a({\bf p}) ) \right\}}|\Psi(0)\rangle
\nonumber\\
&=&\exp{\left\{\int\rd {\bf p}( \alpha({\bf p},t)a^{\dagger}({\bf p})-
\alpha^*({\bf p},t)a({\bf p}) ) \right\}}(a({\bf p'})+\alpha({\bf p'},t))|\Psi(0)\rangle
\nonumber\\
&=&\alpha({\bf p'},t)|\Psi(t)\rangle.
\eea
The final state is a $\varphi$-coherent state with eigenvalue $\alpha$ (cf.\ section 3.4 in \cite{optics}).
This demonstrates that
coherent states in $\varphi$ are associated with $j$-eigenstates in the matter field.
Reduction to a $\varphi$-coherent state
should therefore induce reduction to a $j$-eigenstate in the fermionic sector.

It is tempting to associate $\varphi$ with a gauge field such as the photon field or some proposed graviton field.
The current $j$ would then relate to a conserved charge, e.g.\ electric charge or energy-momentum.
Such charge densities are a natural description of macroscopic observables.

\section{Conclusions}
\label{conc}
The key advance of this paper has been to develop an alternative
model of state reduction in relativistic quantum field theory which does not
suffer from the infinite rates of energy density increase seen in some
previous proposals. We have outlined a model requiring just one extra parameter
in addition to those of standard quantum theories
in order to simultaneously describe the quantum behavior of
individual excitations and the definite behavior of macroscopic
objects.

In our approach, by having no coupling between the classical stochastic field and the
particle creation operator,
we ensure that the evolution equation cannot randomly create particles from the vacuum.
Our model features only a coupling between the stochastic field
and the particle annihilation operator. This is appealing for two
further reasons. First, it leads to a reduction to coherent
states. As coherent states saturate the bound of the Heisenberg
uncertainty relation they make a natural choice as a quantum
counterpart to an idealized classical state. Second, by applying
this mechanism to a bosonic field coupled to a fermionic field, we
can induce state reduction to some charge density basis in the
fermionic sector. The model requires the specification of a preferred set
of space-like hyper-surfaces supporting the time-like state evolution.
This breaks relativistic invariance. However, our perturbative calculations show
no deviation from relativistic invariance to second order in $\lambda$.

The ideas presented in this paper could be applied to the photon field or to a proposed graviton
field in order to see state reduction to a conserved electric charge or
energy-momentum basis in the associated matter fields.
Since the model predicts an energy loss
which could be significant in high-density highly accelerating matter environments,
there may be the possibility of experimental
investigation, e.g., by looking at the decay of high intensity electromagnetic waves
or through the detection of gravitational waves.

\section*{Acknowledgements}
The author wishes to thank Dorje Brody and Philip Pearle for valuable discussions.

\end{document}